# A Graphene-based Hot Electron Transistor

*Sam Vaziri[1,#], Grzegorz Lupina[2,#], Christoph Henkel[1], Anderson D. Smith[1], Mikael Östling[1], Jarek Dabrowski[2], Gunther Lippert[2], Wolfgang Mehr[2], Max C. Lemme[1,3,*]*

[1] KTH Royal Institute of Technology, School of Information and Communication Technology, Isafjordsgatan 22, 16440 Kista, Sweden

[2] IHP, Im Technologiepark 25, 15236 Frankfurt (Oder), Germany

[3] University of Siegen, Hölderlinstr. 3, 57076 Siegen, Germany

**Abstract**

We experimentally demonstrate DC functionality of graphene-based hot electron transistors, which we call Graphene Base Transistors (GBT). The fabrication scheme is potentially compatible with silicon technology and can be carried out at the wafer scale with standard silicon technology. The state of the GBTs can be switched by a potential applied to the transistor base, which is made of graphene. Transfer characteristics of the GBTs show ON/OFF current ratios exceeding $10^4$.

**Keywords**
graphene, transistor, hot electrons, hot carrier transport, tunneling

[#] These authors contributed equally

[*] Corresponding author. Email: max.lemme@uni-siegen.de, Tel: +49-271-740-4035

Graphene has been investigated intensely as a next-generation electronic material since the presence of the field effect was reported in 2004[1]. The absence of a band gap and the resulting high off-state leakage currents prohibit graphene as the channel material in field effect transistors (FETs) for logic applications[2]. While graphene RF analog transistors can exploit the higher carrier mobility[3] and saturation velocity[4], band-to-band tunneling reduces drain current saturation and voltage gain[5-7]. Several alternative graphene device concepts have been proposed that rely on quantum mechanical tunneling. These include graphene / hexagonal boron nitride superlattices[8] or (gated) graphene / semiconductor Schottky barriers[9,10]. Along these lines, we recently proposed a Graphene Base Transistor (GBT)[11], a hot electron transistor (HET)[12-14] with a base contact made of graphene that can potentially deliver superior DC and RF performance[11]. HETs with metallic bases are limited by two mechanisms: carrier scattering and "self-bias crowding" (in-plane voltage drop) in the base material. Optimization becomes a trade-off, since thinning the metal-base reduces scattering, but increases the metal-base resistance and the self-bias crowding[12]. Graphene is thus the ideal material for HET bases due to its ultimate thinness and high conductivity. Theoretical calculations predict that ON/OFF current ratios of over five orders of magnitude and operation up to the THz frequency range can be obtained with GBTs[11]. A schematic cross-section and top-view of a GBT are shown in Figure 1a and b. The graphene base electrode is sandwiched between two insulating dielectrics, which are covered with two electrodes (emitter and collector). The emitter-base insulator (EBI) functions as the tunneling barrier.

In our implementation, the collector is made of metal and the emitter is made of doped silicon. The fabrication process was designed to be largely silicon CMOS technology compatible (see methods section). Here, we are reporting on experimental data of six different devices, labeled "device A" through "device F", including the supporting infomration. A top-view photograph of a GBT is shown in Figure 1c. The specific band structure of the GBTs investigated in this work is shown schematically in Fig. 2 for three relevant cases: (a) the flatband case with no external bias, (b) the OFF-state, where a collector bias is applied and (c) the ON-state with both collector bias and base bias. We note that the work functions, band offsets and bias voltages are drawn to scale based on well-known literature data for the materials used for fabrication, while the layer thicknesses are not to scale. In particular, an n-doped silicon emitter, a thermally grown silicon dioxide ($SiO_2$) EBI tunneling barrier, a graphene base, an atomic layer deposited (ALD) aluminum oxide ($Al_2O_3$) BCI and an evaporated titanium / gold collector contact were used. Without a voltage drop across the EBI (e.g. $V_E = V_B = 0$ V as in Fig. 2a and b), the device is "OFF" regardless of any reasonable positive bias applied to the collector. There should be no current flowing from the emitter to the base or the collector as electrons in the emitter face the high potential barrier of the EBI. In reality, the monoatomic graphene base layer does not fully screen the electrical field generated when a collector bias is applied[8], and there is a slight voltage drop across the EBI as indicated in Fig 2b. When a positive voltage is applied to the base in addition to a finite collector voltage (with $V_B < V_C$), hot electrons will tunnel across the lowered barrier of the EBI from the conduction band of the n-doped silicon to the base through the Fowler-Nordheim mechanism. If all barriers are chosen carefully, these hot electrons are further injected into the base collector insulator conduction band and arrive at the collector contact. Thus, the state of graphene base transistor can be controlled with the potential of the graphene base electrode.

Fig. 3a and b show the wiring and the corresponding measurement of the collector current versus base voltage of a GBT (device A). This measurement is similar to the transfer characteristics (i.e. drain current vs. gate voltage) in standard silicon metal oxide semiconductor (MOS) FETs. In this device, the EBI and the BCI consist of 5 nm $SiO_2$ and 21 nm $Al_2O_3$, respectively. Both base contacts were connected to ensure a more uniform potential distribution across the base. The emitter potential was $V_E = 0$ V and the collector was biased at $V_C = 8$ V. The base voltage was swept from 0 V to 6 V and back to 0 V. At a voltage of $V_{Bth} \approx 4.5$ V the current $I_C$ measured at the collector contact increases rapidly. This is the threshold at which the energy barrier of the EBI is reduced sufficiently to allow Fowler-Nordheim tunneling and, ideally, the electrons have sufficient energy to be injected into the conduction band of the BCI (compare Fig. 2c). It separates the OFF-state form the ON-state and we call $V_{Bth}$ the "threshold voltage" in analogy to conventional MOSFETs. Comparing $I_C$ at graphene base voltages below and above the threshold voltage results in an ON/OFF collector current ratio of > 1000 (compare Fig. 3b). In addition, the dual base voltage sweep reveals that no significant hysteresis is present in the device characteristics. This indicates that charge trapping in the insulators does not play any significant role in the operation. An alternative measurement setup for a different GBT (device B) with identical EBI thickness and 25 nm $Al_2O_3$ BCI is shown in figure 3c. Here, the base and the collector potentials are fixed at $V_B = 0$ V and $V_C = 2$ V, respectively. Instead of the base, the emitter voltage is swept from 0 V to -6 V. The threshold voltage is again reached for a voltage drop across the EBI of 4.5 to 5 V. The inset in Fig. 3d shows the GBT transfer characteristics for the same device, but includes a sharp drop of the collector current at $V_B \approx 6$ V, caused by a hard breakdown of the EBI silicon oxide. As a consequence, the emitter and base were short-circuited and the entire emitter current flows through the base contacts, as the electrons can no longer gain sufficient energy to be injected into the BCI conduction band. Variable temperature measurements (see supporting information) show that in the relevant high-field region, electrons are injected from the emitter to the graphene base mostly by tunneling. The collector current, in contrast, shows some increase with temperature. This indicates that part of the transport through the base collector insulator occurs via a defect-mediated process.

In subsequent measurements, the base and the collector voltage were swept simultaneously. This keeps the electric field across the BCI constant and reduces the stress on the EBI, because it minimizes the exposure time of the device to the maximum electrical field. Here we recall that the collector potential influences also the field in the EBI due to incomplete screening at the graphene base. The band structure for such double sweeps is shown schematically in Fig. 4a. A set of transfer characteristics can be seen in Fig. 4b. This is a different device than those in Fig. 3, with an EBI and a BCI of 5 nm $SiO_2$ and 25 nm $Al_2O_3$, respectively (device C). We use the term "emitter-base voltage" in the figure caption to differentiate from the previous measurements. In the ON-state, the collector current clearly depends on the base-collector voltage difference $V_{BC}$. Figure 4c shows the same data in logarithmic scale. The threshold voltage is similar to the devices in Fig. 3, indicating that the switching mechanism is dominated by the emitter-base tunneling process. These GBTs achieve an ON/OFF collector current ratio of ~$10^3$. Base-collector voltages greater than 6 V lead to an additional increase in the collector current below the threshold voltage. We speculate that this is the onset of additional conduction mechanisms through the $Al_2O_3$ BCI, an undesirable parasitic effect. An additional unexpected collector current increase at low base voltages between $V_B = 0$ V and 1 V is also observed, that is attributed to the charging and discharging of traps in the EBI and/or the BCI.

Figure 4d shows the collector current $I_C$ as a function of the collector voltage, which is the equivalent to output characteristics in conventional MOSFETs. The data is extracted from the previous graphs for different base voltages and a fixed emitter voltage of $V_E = 0$ V. Above the threshold voltage of $V_{Bth} = 4.5$ V, $I_C$ increases rapidly with higher collector voltages. This is in good agreement with our predictions[11]. The collector currents do not saturate, which would be expected, but dielectric breakdown prevents applying sufficiently high collector voltages in this first generation of GBTs. Future BCI materials optimized for band offsets and thickness will extend the window of operation.

The common-base transfer characteristics of a GBT (device D) at $V_{BC} = 6$ V is shown in Fig. 5. The transfer ratio α, defined as the ratio between collector current and emitter current in the ON-state reaches values of up to 6.5% in our devices. While this is comparable to reports on metal-insulator-metal-insulator-metal HETs[12,15,16] (see supporting information), high-energy barriers at both EBI and BCI prevent reaching the full performance potential of these devices. However, structures with low tunneling barriers and optimized thickness of EBI and BCI should enable competitive operation characteristics[11,17].

Finally, we note that the collector currents in the ON state are rather low, too low when addressing potential future applications. One option to improve this is to reduce the thickness and barrier height of the EBI, as a linear decrease in thickness will lead to an exponential increase in the tunneling currents[18]. Another option is to reduce the band offset and the thickness of the BCI, as these will decrease the quantum mechanical reflection at the base-insulator band edge and the scattering rate during transport across the dielectric. An example is shown in Fig. 5, which compares the transfer characteristics of a GBT with a reduced BCI of 16 nm (device E) with the device in Fig. 4. The currents are normalized for size to compensate for different device areas, hence the difference in OFF-state leakage. Apart from the BCI thickness, the fabrication process was identical. A clear increase in $I_C$ can be observed despite a slightly lower $V_{BC}$, along with an increase in the ON/OFF ratio exceeding $10^4$ if the base voltage is extended to 7 V. In comparison with devices in Fig. 3b,d and Fig. 4c, this higher ON/OFF ratio has been achieved with a thinner BCI (16 nm $Al_2O_3$ instead of 25 nm) and a higher emitter-base voltage (6.7 V instead of 6V). This clear dependence of $I_C$ on the BCI thickness and a slight temperature dependence of the BCI currents (not shown) indicate that the devices do not operate in the fully ballistic regime, which is required to obtain high gain. We expect improved performance form lower tunneling barriers and dielectric thicknesses[11,17]. Nevertheless, one should note that small band offsets and thinner BCIs will increase the OFF-state currents through thermionic emission, direct tunneling and Frenkel-Poole tunneling. As a consequence, GBTs will have to be optimized with regards to thickness and materials for BCI and EBI depending on the envisioned application.

In summary, we report the experimental realization of a vertical hot electron transistor that can be switched by a voltage applied to the graphene base. We achieve ON/OFF current ratios exceeding $10^4$ and most of the fabrication process is compatible with CMOS technology. Potential applications for GBTs include high speed RF analog devices like low noise amplifiers or power amplifiers and, if combined with complementary hot hole transistors, logic circuits.

**Methods**

A CMOS compatible process scheme on 200mm silicon (100) substrates was used to fabricate the GBT structures. Neighboring devices were electrically isolated by shallow trench isolation (STI). Trenches were etched into the Si substrate and filled with high density plasma chemical vapor deposited $SiO_2$, followed by chemical mechanical polishing. After a phosphorous implantation step to dope the Si emitter, a 5 nm-$SiO_2$ emitter base insulator (EBI) was grown by thermal oxidation. A photograph of a full processed wafer is shown in the supplementary material. The wafers were then cut into 1 x 1 $cm^2$ chips to facilitate experimental process variations. Commercially available chemical vapor deposited (CVD) graphene were then transferred from their copper substrates similar to the methods described by Li et al. [19] and Lin et al. [20]: A layer of Poly(methylmethacrylate) (PMMA) was spin-deposited to one side of the copper/graphene substrate. Subsequently, the backside graphene was removed in oxygen plasma, and the copper film was selectively etched in a $FeCl_3$ solution. After rinsing in de-ionized water, the PMMA/graphene film was transferred from solution onto the Si chips. PMMA was removed in a two-step wet chemical treatment in Acetone and Chloroform. A forming gas anneal at 350 °C was applied to evaporate residual solvents and polymer. After transfer, the presence and quality of single layer graphene sheets were confirmed by Raman spectroscopy[21]. Considering that the choice of contact metal is not at all limited to Au, we note that the graphene transfer and lift-off are the only process steps not compatible with state-of-the-art silicon technology. The graphene sheet was patterned by photolithography and reactive ion etching. Afterwards, the graphene base contacts of 15 nm Ti / 70 nm Au were deposited by e-beam evaporation in combination with a lift-off technique. The base collector insulator was deposited in two steps. First, a 3 nm Al seed layer was deposited by e-beam evaporation. This thin Al layer transforms completely to aluminum oxide during a subsequent exposure to ambient air. In the second step, Al$_2$O$_3$ was deposited by atomic layer deposition (ALD) using a standard trimethyl-aluminum/water process. The total Al$_2$O$_3$ thickness was confirmed by spectroscopic ellipsometry on bare Si wafers. Finally, a metal stack of 15 nm Ti / 70 nm Au was e-beam evaporated and structured with a lift-off process to form the collector electrode. The devices were electrically characterized as double gate field effect transistors to confirm the presence of graphene (see supplementary information). All measurements were performed at room temperature and ambient air.

ACKNOWLEDGMENTS

The authors thank F. Driussi, P. Palestri and L. Selmi (Univ. Udine) for fruitful discussions. Support from the European Commission through a STREP project (GRADE, No. 317839), an ERC Advanced Investigator Grant (OSIRIS, No. 228229) and an ERC Starting Grant (InteGraDe, No. 307311) as well as the German Research Foundation (DFG, LE 2440/1-1) is gratefully acknowledged.

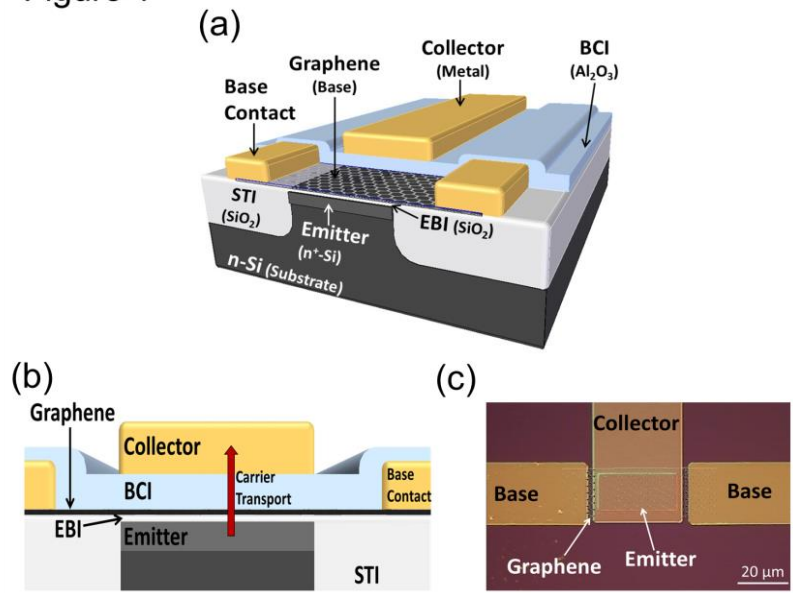

**Figure 1:** (a) Schematic layout of the three terminal graphene base transistor (GBT). The emitter is formed by the doped Si substrate. The graphene base is transferred on top of the emitter after forming a thin emitter-base insulator (EBI). The graphene base is contacted and a collector-base insulator (BCI) is deposited on top of the graphene base before depositing the metal collector. (b) Cross-section of a GBT. During device operation, hot carriers are injected from the emitter across the EBI and the graphene base into the collector, as indicated by the red arrow. (c) Top view optical micrograph of a GBT with two base contacts. A cartoon of the graphene base has been added for clarity.

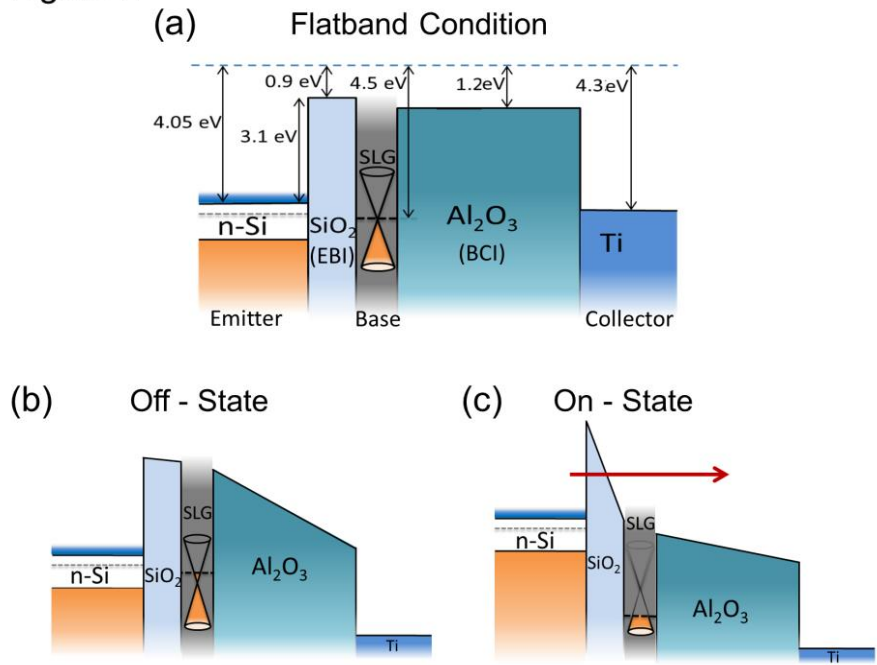

**Figure 2:** Schematic band diagram of a GBT in different modes of operation (drawn to scale on the energy axis). The materials are identical to the ones used in the experiments. The graphene layer is assumed to be undoped, which is most likely different from the experiment. However, the results are not generally affected by the doping level. (a) The band alignment under flat band condition. (b) For finite collector voltages the device is in the OFF-state. A slight influence on the EBI field is shown to take into account incomplete screening of the collector field by the graphene base[8]. (c) Increasing the base voltage to more positive voltages switches the device to the ON-state. The effective tunneling barrier of the EBI is reduced to enable Fowler-Nordheim tunneling, ballistic transport across the graphene, and injection of hot electrons into the BCI conduction band.

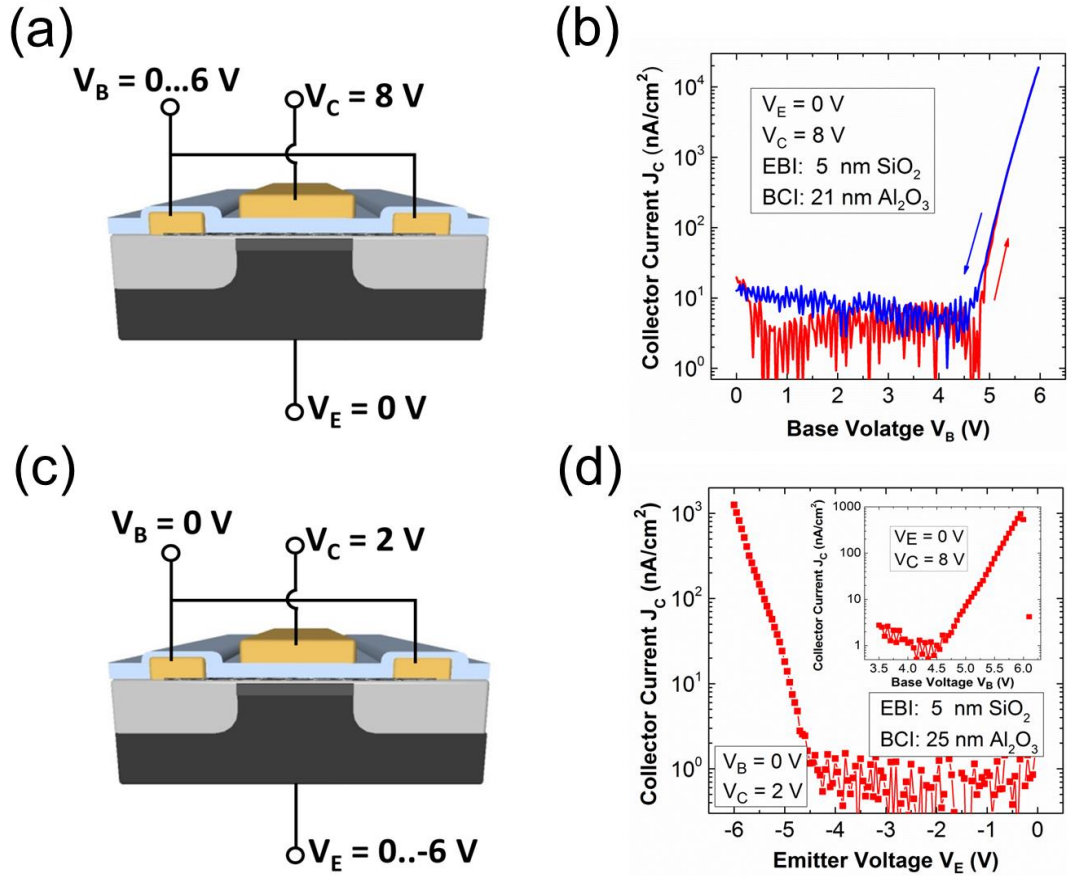

**Figure 3:** (a) Schematic cross section of the GBT wiring setup for a base voltage sweep. (b) Transfer characteristics of a GBT (device A). The graphene base voltage is swept from 0 to 6 V (red) and back to 0V (blue) while biasing the emitter and the collector at 0 and 8 V, respectively. The collector current $I_C$ is monitored. An ON/OFF collector current ratio of $10^3$ is achieved. (c) Schematic cross section of the GBT wiring setup for an emitter voltage sweep. (d) The emitter voltage $V_E$ is swept from 0 to -6 V while biasing the base and the collector at 0 and 2 V, respectively (device B). Inset: Transfer characteristics for the same device, including breakdown at $V_B = 6$ V.

# Figure 4

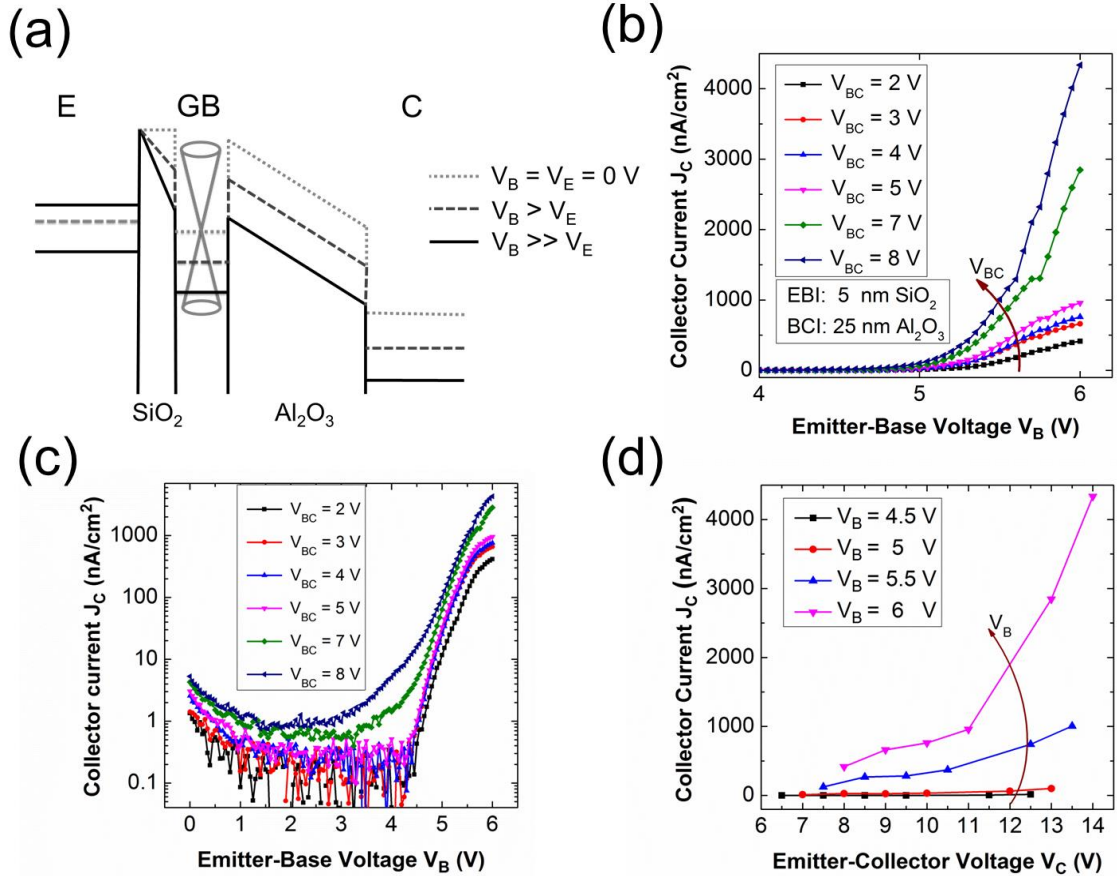

**Figure 4:** (a) Idealized schematic band diagram during double sweep operation. The graphene base voltage and the collector voltage are kept at a certain fixed voltage difference. The injection of hot electrons from the n-doped Si emitter is controlled entirely by the EBI field. (b) Transfer characteristics for a fixed base collector bias $V_{BC}$ and a base voltage sweep from $V_B = 4$ to 6 V. The emitter voltage is kept at 0 V (device C). (c) Logarithmic scale of the transfer characteristics with an ON/OFF-ratio $> 10^3$. (d) Output characteristics of the GBT for various base voltages $V_B$ extracted from the measurements shown in 4b and c.

Figure 5

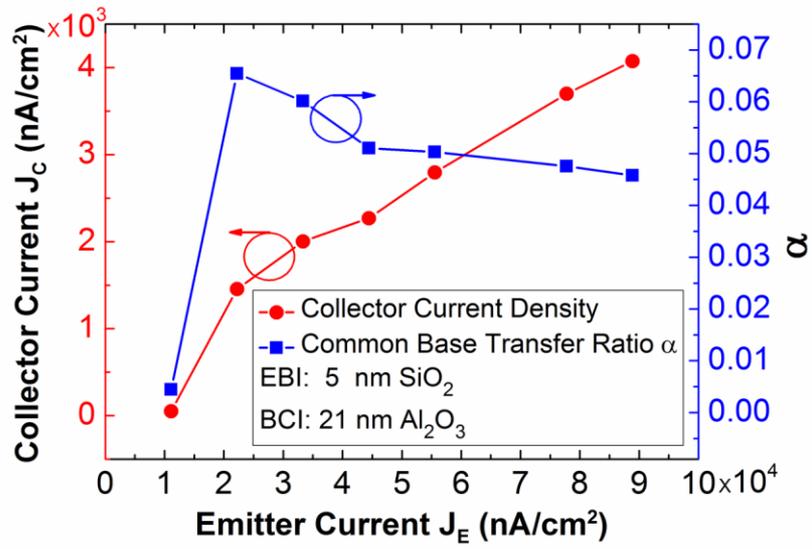

**Figure 5:** Common base transfer characteristics of a GBT at $V_{BC}$ = 6 V. A maximum common base gain up to 6.5% is achieved (device D).

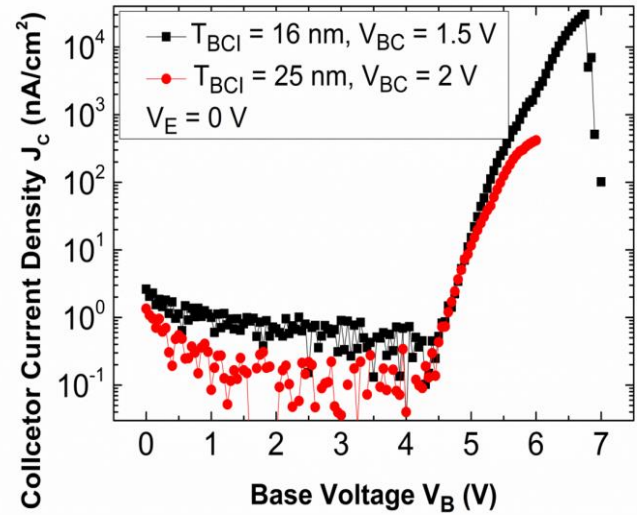

**Figure 6:** Transfer characteristics of a GBT with reduced BCI thickness of 16 nm at a constant base collector voltage difference of $V_{BC}$ = 1.5 V (black squares, device E). A comparison with the device from Fig. 4 with a BCI thickness of 25 nm and $V_{BC}$ = 2 V (red dots) shows a drastic increase in ON-current density and ON/OFF ratio of approximately $5 \times 10^4$. The currents were normalized for size because the devices have different active areas. The 16 nm BCI broke down at $V_B \approx 6.7$ V.